\documentclass{llncs}
\usepackage{makeidx}  

\usepackage{color}
\usepackage{ulem}
\usepackage{times,bm}
\usepackage{multicol}
\usepackage{graphicx}
\usepackage{epstopdf}
\usepackage{amsmath, amssymb}
\usepackage{type1cm}
\usepackage{mathtools}
\usepackage{multirow}
\usepackage{wrapfig}
\usepackage{here}
\usepackage{geometry}
\begin{document}
\frontmatter          
\mainmatter              
\title{
Semi-supervised Approach to Soft Sensor Modeling for Fault Detection in Industrial Systems with Multiple Operation Modes
}
\author{Shun Takeuchi \and Takuya Nishino
\and Takahiro Saito \and Isamu Watanabe}
\authorrunning{Shun TAKEUCHI et al.} 
\institute{
Artificial Intelligence Research Center,
Knowledge Information Processing Laboratory,
Fujitsu Laboratories Ltd.,
Japan,\\
\email{takeuchi.shun@jp.fujitsu.com}}

\maketitle              

\begin{abstract}
In industrial systems, certain process variables that need to be monitored for detecting faults are often difficult or impossible to measure. Soft sensor techniques are widely used to estimate such difficult-to-measure process variables from easy-to-measure ones. Soft sensor modeling requires training datasets including the information of various states such as operation modes, but the fault dataset with the target variable is insufficient as the training dataset. This paper describes a semi-supervised approach to soft sensor modeling to incorporate an incomplete dataset without the target variable in the training dataset. To incorporate the incomplete dataset, we consider the properties of processes at transition points between operation modes in the system. The regression coefficients of the operation modes are estimated under constraint conditions obtained from the information on the mode transitions. In a case study, this constrained soft sensor modeling was used to predict refrigerant leaks in air-conditioning systems with heating and cooling operation modes. The results show that this modeling method is promising for soft sensors in a system with multiple operation modes.
\keywords{Fault detection, Soft sensor, Semi-supervised learning, Constrained optimization}
\end{abstract}
\section{Introduction}
Process monitoring is an essential element of operating industrial processing equipment. Process variables, such as temperature, pressure, and flow rate, are measured by a large number of instrumented sensors in industrial systems. Process monitoring is needed for controlling processes and detecting faults, and it ensures stable operation of these systems \cite{chiang2000fault,dunia1998joint,ge2012multivariate,yang2015}.
The requirement for fault detection is to measure the target process variables online. However, certain important process variables, such as product composition in distillation columns, are difficult or impossible to measure online. Such difficulties may stem from particular technical problems, time-consuming analyses, and/or the high cost of measuring devices. Soft sensor (or virtual sensor) techniques have been studied as ways of monitoring such difficult-to-measure process variables.

Soft sensor techniques have been developed over the course of the last two decades \cite{fortuna2007soft,kadlec2009data}. Soft sensors monitor target processes by modeling the relation between the easy-to-measure (or input) variables and the difficult-to-measure (or output) variable. The input variables are the processes to be measured by the instrumented sensors (i.e., hardware or "hard" sensors) in the systems. By applying the measured input variables to the soft sensor model, it becomes possible to estimate the output variable online. Soft sensors are implemented with the purpose of improving the quality and efficiency of industrial systems.

There are two major classes of soft sensors: model-driven and data-driven \cite{kadlec2009data,kalos2005data}. A model-driven soft sensor is a first principles model (FPM) based on physical properties. Therefore, it usually focuses on describing the ideal steady state of the process; this is a drawback when it comes to applying it to real-world industrial processes. On the other hand, thanks to progress in the development of machine learning techniques, the data-driven soft sensors have been used to the process monitoring
\cite{lin2007systematic,liu2010developing,Serpas2013,tian2016outliers}.

Industrial systems often have multiple operation modes. For instance, industrial plants usually have a number of production processes, and they switch modes depending on the purpose of those processes. The dynamics controlling the behavior of the process variables in each operation mode are different. Therefore, the statistical properties of the process variables are also different \cite{sari2014data}. The prediction accuracy of soft sensors is reduced by changes in the statistical properties of the process variables. This problem is called degradation of a soft sensor model \cite{fortuna2007soft,kadlec2011review,kaneko2013classification}.

In soft sensor based fault detection, one detects process faults by monitoring the time variation of the estimated output variable. However, it is difficult to appropriately detect the process faults in a situation in which the prediction accuracy of each operation mode may deteriorate. That is, when the operation mode changes, the error of the estimated output variable changes even if process faults do not occur, and this results in incorrect assessments of the process faults. In this paper, we propose a soft sensor modeling method for fault detection in industrial systems with multiple operation modes.

The paper is organized as follows. The next section presents the soft sensor modeling with multiple operation modes and defines the problem. Section 3 describes our method. Section 4 illustrates the results of an experiment that applied the method to real process data of an industrial system. The final section is devoted to concluding remarks.

\section{Soft Sensor Modeling with Multiple Operation Modes}
For the model of data-driven soft sensor in industrial systems with multiple operation modes, adaptive soft sensors have been developed \cite{garcia2009fault,qi2011enhanced,yuan2014soft}.
The adaptive model selects an appropriate training dataset for the target mode from a database. 
There are several types of adaptive model. For example, the Just-In-Time (JIT) models are constructed with only data close to a query sample or with all data having weights according to similarity with a query sample \cite{fujiwara2009soft,kadlec2011review}.
By extracting an appropriate training dataset in this manner, the prediction accuracy of the model is improved. Note that the basic idea of these models is that the training dataset is divided up by operation mode and appropriate models are constructed for each operation mode.

The foundation of the soft sensor model is multiple linear regression analysis \cite{bishop2006pattern}. The formula of the model is
\begin{eqnarray}
y_n^{(m)} = \bm{x}_n^{(m)} \bm{w}^{(m)} + \epsilon^{(m)},
\end{eqnarray}
where $y \in \mathbb{R}$ is the output variable, $\bm{x} \in \mathbb{R}^M$ are the input variables, $\bm{w}$ is the regression coefficient, $\epsilon$ is the residual, $M$ is the dimensionality of the samples, and $n =1,2,\ldots,N$ is the number of observations. Here, we represent the identifier of the operation mode with the superscript $m$. One can obtain the regression coefficient by using the least squares method:
\begin{equation}
\begin{aligned}
& \underset{w}{\text{minimize}}
& & \sum_{n=1}^N \left( \bm{x}_n^{(m)} \bm{w}^{(m)} - y_n^{(m)} \right)^2.
\end{aligned}
\end{equation}
By separating the dataset into subsets for each operation mode, an appropriate regression coefficient is obtained for the particular operation. When the amount of training data for each operation mode is sufficient for representing the properties of the output variables, errors do not arise at mode transitions. Thus, this model can accurately estimate the time variation of the output variable and evaluate process faults.

However, one of the problems of soft sensor development for fault detection is insufficient training data \cite{lou2002comparison}. Since the output variable is difficult or impossible to measure, one needs to obtain datasets with input and output variables by conducting off-line fault tests in a laboratory. On the other hand, it is known that data of industrial systems often "drift" as a result of changing environmental conditions (e.g., changes in the weather) \cite{kadlec2009data}. Since drift is caused the variance of the data properties, a number of training datasets is needed. However, it is not easy to perform time-consuming fault tests under various conditions.

\section{Proposed Model}
\subsection{Overview}
Our strategy for solving the problem of insufficient training data is to treat the dataset without the output variable as the training dataset for the soft sensor modeling. For simplicity, we will consider here a soft sensor model of an industrial system with only two operation modes. Suppose that the dataset of these two modes has been divided up by using some approach such as 
expert knowledge.

Now let us define the two datasets shown in Table \ref{table:training}. The complete dataset includes the input variables and the output variable in each operation mode. These are used as the training dataset in traditional soft sensor modelings such as the datasets of off-line fault tests conducted in a laboratory. The incomplete dataset includes only the input variables in each operation mode. Such datasets are acquired under the conditions in which it would be difficult or impossible to measure the output variable, such as past real process data of the system. Traditional fault diagnosis by experts (e.g., monitoring of easy-to-measure single process) is used to make empirical assessments of fault occurrences. It is possible to obtain the datasets of a system that was qualitatively evaluated as being faulty, although the value and time variation of the target output variable remain unknown. We pointed out the problem of incomplete data in the previous section. The purpose of this study is to overcome this problem by making it possible to use using the incomplete data.
\begin{table*}[t]
 \begin{center}
 \caption{Training datasets of the proposed method.}
 \label{table:training}
  \begin{tabular}{l|c|c|c}
   \hline
    Dataset & Operation modes & Input variables & Output variable \\
   \hline
    Complete dataset & $m=1, 2$  & $\surd$ & $\surd$ \\
   \hline
    Incomplete dataset & $m=1, 2$ & $\surd$ & No \\
   \hline
  \end{tabular}
\end{center}
\end{table*}

The incomplete dataset without the output variable cannot usually be used as the training dataset for regression analysis. Here, let us focus on data at the transition point between operation modes. The properties of industrial processes often change when switching operation modes. On the other hand, faults that occur due to external factors occur independently of operating mode changes. That is, the process variables relevant to the faults do not increase or decrease rapidly as a result of mode transitions. Therefore, we evaluated the regression coefficients of each operation mode under the condition that the difference between the estimated output variables of the operation modes does not become large.

\subsection{Constrained regression model}
We developed a model based on constrained least squares \cite{lange2010numerical,van1996matrix}. As mentioned in the previous section, our method is based on multiple linear regression modeling in each operation mode. The estimated output variables in each operation mode are given by $\hat{y}_n^{(m)} = \bm{x}_n^{(m)} \bm{w}^{(m)}$. Operation mode 1 switches to 2 at the elements $n = i$ and $n = j$. At the transition point of the operation modes, the condition imposed on the estimated variables is expressed as follows:
\begin{eqnarray}
| \hat{y}^{(2)}_{j} - \hat{y}^{(1)}_{i} | &\le& c.
\end{eqnarray}
The constant $c \ge 0$ is the constraint parameter, which it would be near zero in this context. As a result, the constraint condition for the regression model is defined by
\begin{eqnarray}
g (\bm{W}) =
{\bf X} \, {\bm W} + c {\bm 1} 
=
\begin{pmatrix*}[r]
 \bm{x}^{(1)}_{i} & -\bm{x}^{(2)}_{j} \\
-\bm{x}^{(1)}_{i} &  \bm{x}^{(2)}_{j} 
\end{pmatrix*}
\begin{pmatrix*}[r]
 \bm{w}^{(1)}  \\
 \bm{w}^{(2)} 
\end{pmatrix*} +
c {\bm 1} \ge 0,
\end{eqnarray}
where $\bm{W} = \begin{bmatrix} \bm{w}^{(1)} & \bm{w}^{(2)} \end{bmatrix}^{\top}$ are the regression coefficients for the operation modes, $\bm{X}$ is the matrix of input variables at the transition point, and ${\bm 1}$ is a unit vector. Note that the constraint condition does not include the values of the output variable.

In this constraint optimization problem, we need to minimize the objective function, $f(\bm{w}) = \sum_{n} \left( \bm{x}_n \bm{w} - y_n \right)^2$, for each operation mode. This problem is called multi-objective optimization \cite{bandaru2017data}. Since the industrial processes of one mode are independent of those of another mode, we thus use a single objective function which linearly combines the two objective functions, i.e., 
\begin{eqnarray}
f(\bm{W}) = \sum_{n=1}^{N^{(1)}} \left( \bm{x}_n^{(1)} \bm{w}^{(1)} - y_n^{(1)} \right)^2 
             + \sum_{n=1}^{N^{(2)}}  \left( \bm{x}_n^{(2)} \bm{w}^{(2)} - y_n^{(2)} \right)^2.
\end{eqnarray}
Consequently, we obtain an equation of a constrained least squares for solving the regression coefficients $\bm{W}$, as follows:
\begin{equation}
\begin{aligned}
& \underset{W}{\text{minimize}}
& & f(\bm{W}) 
& \text{subject to}
& & g_\ell (\bm{W}) \geq 0,
& & \ell =1,\ldots,p,
\end{aligned}
\end{equation}
where $p$ is the number of transition points.

This equation means that the more transition points of operation modes there are in the incomplete dataset, the more the constraints there are on the complete dataset. In other words, the number of training datasets increases. This method is applicable to industrial systems with multiple operation modes. By taking into account the mode transitions of industrial systems and the physical properties of the output variable, it becomes possible to incorporate an unlabelled incomplete dataset in the training dataset. Such a semi-supervised learning approach can solve the problem of insufficient datasets in fault detection modeling.

\subsection{Procedure of proposed method}
The proposed method is summarized as follows:
\begin{description}
\item[Step 1:] select incomplete datasets and the elements at the transition points of operation modes;
\item[Step 2:] calculate the estimated output variables in the operation modes;
\item[Step 3:] define the constraint condition under which the difference between the estimated values of the modes do not become large with a constraint parameter;
\item[Step 4:] under that condition, evaluate the regression coefficients of each operation mode by using complete datasets.
\end{description}

\section{Case Study}
\subsection{Data Description and preprocessing}
This section presents the development of soft sensor model based on the proposed method and real process data of an air-conditioning system. The air-conditioning system has generally two operation modes, heating and cooling, and makes adaptive transitions between modes in accordance with changes to the environment. Several hard sensors measure processes such as temperature, pressure of the refrigerant, and rotation frequency of the compressor, and are used for control and monitoring \cite{kim2010development,wang2004ahu,xiao2006diagnostic}. On the other hand, the refrigerant, the amount of which is important for stable operation, is technically difficult or impossible to measure. A decrease in the amount (i.e., a refrigerant leak) is a fault that occurs through fatigue and/or corrosion of the refrigerant piping, and it is generally independent of changes in operation mode. We constructed a soft sensor model for refrigerant leaks.
\begin{table*}[t]
 \begin{center}
 \caption{Datasets of air-conditioning system.}
 \label{table:expdata}
  \begin{tabular}{c|c|c|c|c}
   \hline
    Dataset & Description & Operation mode & Input variables & Output variable \\
   \hline
    A & Fault test data & Heating & $\surd$ & $\surd$ \\
   \hline
    B & Fault test data & Cooling & $\surd$ & $\surd$\\
   \hline
    C & Real process data & Heating, cooling & $\surd$ & No \\
   \hline
    D & Real process data & Heating, cooling & $\surd$ & No \\
   \hline
  \end{tabular}
\end{center}
\end{table*}

We used four datasets to validate our model (Table \ref{table:expdata}). Dataset~A was the dataset of the off-line fault test for refrigerant leaks conducted in a laboratory. The fault test was conducted in the heating mode, and the dataset consists of input variables and the output variable (degree of leakage).  Dataset~B was similar to dataset~A, but its fault test was conducted in cooling mode. Dataset~C was the real process data, which consisted of only input variables in both operation modes. Dataset~D was the same as dataset~C, but it was taken from another unit of the system. An empirical assessment made by experts indicated that the systems corresponding to datasets C and D had refrigerant leaks. These systems were stopped for repair on April 14, 2016 (dataset C) and December 24, 2015 (dataset~D), and they restarted operation on May 31, 2016 (dataset~C) and April 22, 2016 (dataset~D).

The units and range of the process variables were not uniform because they had meanings in the original domain of the system (e.g., temperature and pressure). To obtain an accurate predictive model, we applied Z-score normalization to the process variables to give them a mean equal to 0 and a variance of 1 \cite{han2011data}.

Process variables are usually correlated with each other through the specific operation processes of the system. In our case, there was a high degree of collinearity (redundancy) among the variables, which would reduce the accuracy of the model prediction \cite{kadlec2009data}. Here, we applied principal component analysis (PCA) to the input variables in the training dataset, which avoids the effect of collinearity. PCA produced two matrixes, $\bm{x}_n^{(m)} = \bm{t}_n^{(m)} {\bf P}^{(m) \top}$, where $\bm{t}_n \in \mathbb{R}^K$ are scores, and ${\bf P} \in \mathbb{R}^{M \times K}$ is the loading. We used the scores in the model instead of input variables. This approach is called principal component regression (PCR) \cite{jolliffe2002principal}. In this study, the number of principal component $K$ was empirically determined to cover 80\% of the variance \cite{valle1999selection}.
\begin{table*}[t]
 \begin{center}
 \caption{Combination of training and test datasets.}
 \label{table:comb}
  \begin{tabular}{c|c|c|c|c|c|c}
   \hline
   \multirow{3}{*}{Method} & \multicolumn{3}{|c|}{\multirow{2}{*}{Training dataset}} & \multicolumn{3}{|c}{Test dataset}  \\
   \cline{5-7}
   & \multicolumn{3}{|c|}{} & \multicolumn{2}{|c|}{Test I} &  \multicolumn{1}{|c}{\multirow{2}{*}{Test II}} \\
   \cline{2-6}
   & Heating & Cooling & Constraint & Heating & Cooling &  \\
   \hline
   MPCR & A & B & -- & A & B & C, D \\
   \hline
   SPCR & \multicolumn{2}{c|}{A + B} & -- & A & B & C, D \\
   \hline
   \multirow{2}{*}{CPCR} & A & B & C & A & B & D \\
   \cline{2-7}
   & A & B & D & A & B & C \\
   \hline
  \end{tabular}
\end{center}
\end{table*}

Table \ref{table:comb} lists the combinations of training and test datasets. Three methods were compared. The multiple PCR (MPCR) method constructs a soft sensor model based on multiple linear regression analysis for each operation mode. If the training dataset is sufficient for modeling the behavior of the output variable, it is expected to be accurate. The single PCR (SPCR) method constructs the model without considering the property changes due to mode transitions. The proposed constrained PCR (CPCR) method had two patterns, whose difference depended on the dataset for the constraint condition (dataset C or D). We used the dataset that was not used as the training dataset in Test II. We set $c=0$ and selected the elements of the transition points in the time interval between these operation modes, $| t^{(2)}_{j} - t^{(1)}_{i} | \leq 1$ [day], to exclude the long resting state such as the repair.

We performed two validation tests that quantitatively evaluated the prediction performance. In Test I, the coefficient of determination index $r^2$ and the root mean square error (RMSE) were calculated for datasets A and B. In Test II, we calculated the average of the differences between the estimated the output variables on either side of the transition point of the operation modes transition, $\Delta \hat{y} = | \hat{y}^{\rm (Heat)} - \hat{y}^{\rm (Cool)} | $.

\subsection{Results}
The results of Test I and II are shown in Table \ref{table:fit}.
The coefficient of determination index $r^2$ denotes the correlation between the prediction and the observed value; the higher $r^2$ is, the better the result becomes. RMSE describes the variance of the predicted error; the smaller RMSE is, the more accurate the model becomes. The CPCR method had a lower $r^2$ and higher RMSE compared with the MPCR method, whose model was more fitted to the training dataset. 
On the other hand, the CPCR method had better $r^2$ and RMSE compared with the SPCR method. 

The results of Test II shows the mean values of the difference between the estimated variables at the transition points of the operation mode, $\Delta \hat{y}_C$ and $\Delta \hat{y}_D$. The subscripts C and D mean the applied test dataset. This test evaluated the generalization performance, because the dataset was different from the training dataset. The appropriate model had a small $\Delta \hat{y}$, because the estimated degree of refrigerant leakage did not change much at the transition point. The proposed method thus had better generalization performance compared with the other methods.
\begin{table*}[b]
 \begin{center}
 \caption{Model comparison for validation test.}
 \label{table:fit}
  \begin{tabular}{c|c|c|c|c|c|c}
   \hline
   \multirow{3}{*}{Method} & \multicolumn{4}{|c|}{Test I} & \multicolumn{2}{|c}{Test II} \\
   \cline{2-7}
   & \multicolumn{2}{c|}{$r^2$} & \multicolumn{2}{|c|}{RMSE} & \multirow{2}{*}{$\Delta \hat{y}_{\rm C}$} & \multirow{2}{*}{$\Delta \hat{y}_{\rm D}$} \\   
   \cline{2-5}
   & Heating & Cooling & Heating & Cooling && \\
   \hline
   MPCR & 0.905 & 0.960 & 0.308 & 0.200 & 0.557 & 0.734 \\
   \hline
   SPCR & 0.836 & 0.836 & 0.405 & 0.405 & 0.994 & 0.641 \\
   \hline
   \multirow{2}{*}{CPCR} & 0.854 & 0.880 & 0.382 & 0.347 & -- & 0.209 \\
   \cline{2-7}
    & 0.879 & 0.894 & 0.348 & 0.325 & 0.104 & -- \\
   \hline
  \end{tabular}
\end{center}
\end{table*}

\begin{figure*}[t]
\begin{center}
\includegraphics[width=120mm]{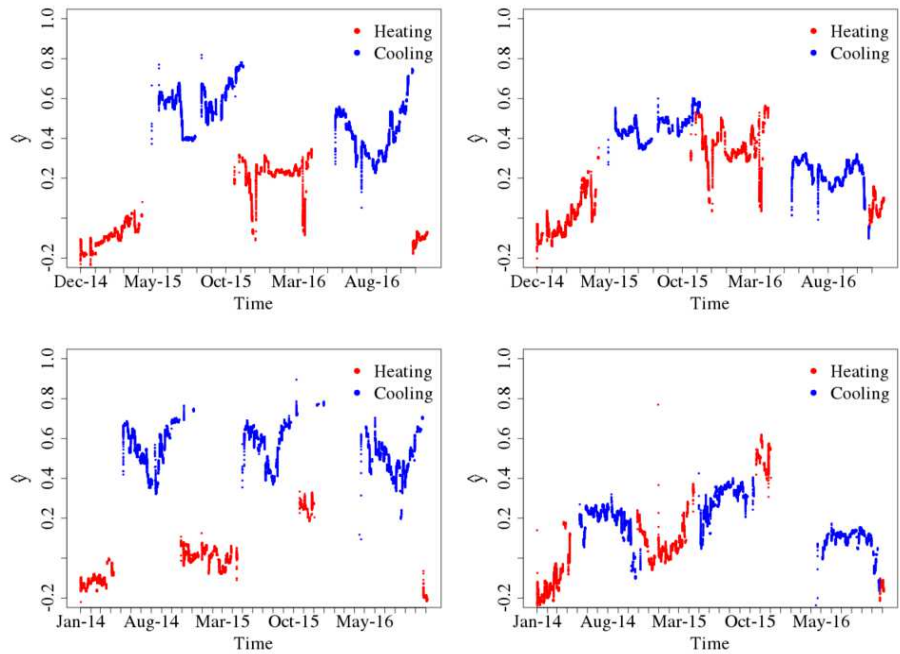}
\caption{
Time series of estimated output variable with operating mode change for MPCR (left panels) and CPCR (right panels) methods applied to datasets C (upper panels) and D (bottom panels).
}
\label{fig:timeseries}
\end{center}
\end{figure*}

Figure~\ref{fig:timeseries} shows the time series of the estimated output variable for an operating mode change. The left panels are the results that applied the MPCR method to dataset~C (the top panel) and D (the bottom panel). The right panels are the results of the CPCR method in which the datasets for the constraint were dataset~D (the top panel) and C (the bottom panel). The red and blue dots indicate results for the heating and cooling modes, respectively. When the operation mode changes, the performance of the MPCR method deteriorates in the left panels. The time series obtained by the MPCR method shows invalid behavior in which the series drastically changes at the transition point of the operation mode. These results provide the wrong diagnosis of a refrigerant leak at the point. On the other hand, the right panels show that the CPCR method can monitor the time variation without predicting an unlikely increase or decrease in the amount of refrigerant. The results of the method are physically realistic, because the time series smoothly change at the point. We thus confirmed the improvement had by the addition of the sample of the constraint.

\section{Conclusions}
We proposed a new semi-supervised approach to soft sensor modeling for fault detection. To alleviate the problem of insufficient data in the input and output feature set of the target process fault, the method uses a real process dataset without the output variable and takes the properties of mode transitions into account. A case study showed that it alleviates the problem of insufficient data. Although we studied a system with only two modes, we can easily extend our method so that it can work with a system having several modes.

\section*{Acknowledgements}
We would like to thank the Office of Air Conditioner Products Development of Fujitsu General Limited for providing us with the air-conditioning system datasets.

\bibliography{ref}

\end{document}